\documentstyle[12pt,aps,epsf]{revtex}
\def\er#1#2{\relax\ifmmode{}^{+#1}_{-#2}\else$^{+#1}_{-#2}$\fi}
\newcommand{\be}{\begin{equation}}
\newcommand{\bea}{\begin{eqnarray}}
\newcommand{\ee}{\end{equation}}
\newcommand{\eea}{\end{eqnarray}}

\newcommand{\krig}[1]{\stackrel{\circ}{#1}}
\def\({\Big(}
\def\){\Big)}

\def\slashchar#1{\setbox0=\hbox{$#1$}
   \dimen0=\wd0 \setbox1=\hbox{/} \dimen1=\wd1
   \ifdim\dimen0>\dimen1 \rlap{\hbox to \dimen0{\hfil/\hfil}} #1
   \else  \rlap{\hbox to \dimen1{\hfil$#1$\hfil}} / \fi}

\begin{document}
\draft
\tighten
\def\footnoterule{\kern-3pt \hrule width\hsize \kern3pt}

\title{
Unitarized Heavy Baryon Chiral Perturbation Theory
and the $\Delta$ Resonance in $\pi N $ Scattering\footnote{Work supported 
by the Junta de Andaluc\'\i a and DGES PB98 1367} 
}
\author{
J. Nieves\footnote{email:jmnieves@ugr.es}
 and E. Ruiz Arriola\footnote{email:earriola@ugr.es}
} 

\address{
{~} \\
Departamento de F\'{\i}sica Moderna \\
Universidad de Granada \\
E-18071 Granada, Spain
}
\date{\today}
\maketitle

\thispagestyle{empty}

\begin{abstract}
Based on the recent phenomenological analysis done by 
M. Mojzis, Eur. Phys. Jour. {\bf C2} (1998) 182, of $\pi N $ 
scattering within the framework of 
Heavy Baryon Chiral Perturbation Theory we predict, after a suitable 
unitarization of the amplitude, the $\Delta$ resonance at $\sqrt{s} = 
M_\Delta = 1238 \er{22}{18}{\rm MeV} $ and with a width of $ \Gamma_\Delta = 
150 \er{43}{31} {\rm MeV} $ in excellent agreement with the experimental 
numbers. The error bars reflect only the uncertainties in the low energy 
parameters as determined from an  extrapolation to threshold of experimental 
data. We also describe satisfactorily, within errors, the phase shifts up to $
\sqrt{s} \sim 1300 {\rm MeV} $.

\end{abstract}

%\vspace*{\fill}\begin{center}Submitted to: {\it }\end{center}

\vspace*{1cm}
\centerline{\it PACS: 11.10.St;11.30.Rd; 11.80.Et; 13.75.Lb;
14.40.Cs; 14.40.Aq\\}
\vspace*{1cm}
\centerline{\it Keywords: 
Chiral Perturbation Theory, Unitarity, $\pi N$-Scattering, }
\centerline{\it $\Delta$ Resonance, Heavy baryon, Partial waves.}

%\vspace*{\fill}
\newpage
%\footnotesize
%\twocolumn
\setcounter{page}{1}

\section{Introduction}
The $\Delta$ resonance plays a prominent role in intermediate
energy nuclear physics \cite{EW88}. Experimentally, it can be seen
as a clear resonance of  the partial amplitude $ f_{2I,2J}^l ( s )$, 
in the $ l_{2I\, 2J} = P_{33} $ channel in $\pi N $ scattering
experiments, at a Center of Mass (CM) energy $ \sqrt{s}=M_\Delta =
1232 {\rm MeV} $ and with a width $ \Gamma_\Delta = 120 {\rm MeV}$
\cite{AS95}.  In this paper we are concerned with the possibility of
generating this  resonance using known information at threshold,
chiral symmetry constraints  and exact unitarization of the
amplitude.  

The practical description of resonances requires, in
general, the use of some unitarization method since exactly at the
resonance energy the amplitude becomes purely imaginary and takes
the maximum value allowed by unitarity \cite{MS72}. On the other
hand, most unitarization methods are based on  perturbation theory
and hence require a resummation of some perturbative  truncated
expansion of the full amplitude. This obviously requires the 
existence of an energy region where the unitarized amplitude
approximately  coincides with the perturbatively expanded one. This
region is typically  close to threshold, since exactly at that point
the amplitude is purely real,  and unitarity sets no constraints on
it. In other words, the amplitude is  justifiedly unitarizable if
perturbation theory works reasonably well somewhere, for instance at
threshold. If this is not the case there is nothing much one can do
about it; one can formally unitarize the amplitude but the
corresponding predictions at threshold will be in general very
different from those obtained in perturbation theory. 

The modern way to effectively incorporate chiral symmetry and departures
from it, is by means of Chiral Perturbation Theory
(ChPT)\cite{gl84}. Being  a perturbative Lagrangian approach it
preserves exact crossing, perturbative renormalization and unitarity
and allows for a bookkeeping ordering of Chiral symmetry
breaking. All detailed information on higher energies or underlying
microscopic detailed dynamics is effectively encoded in some low
energy coefficients which, for the time being, can only be
determined experimentally. For processes involving only pseudoscalar
mesons the expansion parameter is taken to be $ p^2/(4\pi f)^2 $
with $p$ the four momentum of the pseudoscalar meson and $f$ the
weak pion decay constant \cite{gl84}. Thus, this expansion works
better in the region around  threshold. Away from threshold,
however, the expansion breaks down and the  violation of unitarity
becomes increasingly strong. Recently, the use of  unitarization
methods for pseudoscalar mesons has been shown to work, i.e.  the
{\it predicted} unitarized amplitudes reproduce the threshold region,
and provide definite theoretical central values and error estimates
of the  phase-shifts away from threshold and up to about 1 GeV
\cite{EJ99}.

The extension of ChPT to include both mesons and baryons as explicit
degrees of freedom becomes possible if fermions are treated as heavy
particles but in a covariant framework \cite{IW89}, yielding to the so
called Heavy-Baryon Chiral Perturbation Theory (HBChPT)
\cite{JM91,BK92,BKM95}. Here, the expansion to order $N=1,2,3, \dots$
is written in terms of a string of terms of the form $ e^N /( f^{2l}
M^{N+1-2l} ) $, with $l=1, \dots , [(N+1)/2] $. Here, $f$ and $M$ are
the weak pion decay constant and baryon mass respectively. The
quantity $e$ is a generic parameter with dimensions of energy built up
in terms of the pseudoscalar momenta and the velocity $v^\mu$ ($ v^2
=1$ ) and off-shellness $ k $ of the baryons, being the latter defined
as usual through the equation $p_B=\krig{M}v +k $, with $p_B$ the
baryon four momentum and $\krig{M}$ the baryon mass up to corrections
generated by higher orders in the HBChPT expansion. Practical
calculations \cite{BKM97,Mo98,FMS98} show, however, that the
convergence rate of such an expansion may not be as good as it was in
the purely mesonic case. Even at threshold, where the finite baryon
mass effects should be minimal, there appear sizeable corrections to
the scattering lengths for the lowest partial waves. The situation
obviously gets worse as one departs from the threshold region. In the
particular case of $\pi N$ scattering, a systematic calculation has
been done, so far, up to third order \cite{Mo98,FMS98}, i.e. up to and
including terms of order $1/f^2 $ (first order), $ 1/(f^2M)$ (second
order) and $1/f^4$ and $1/(f^2 M^2)$ (third order). A fit to threshold
properties allows to extract the low energy constants, with the result
that the second order contribution turns out to be larger than the
first order one!\footnote{Both Ref.~\cite{Mo98} and Ref.~\cite{FMS98}
make a fit of the low energy contribution either by extrapolation of
close to threshold data to threshold or by extrapolation of the theory
to the close to threshold region respectively, but only the first one
displays the contributions of the several orders explicitly, so we
will mainly refer to Ref.~\cite{Mo98} in this regard. }. For instance,
in the $P_{33}$ channel Mojzis \cite{Mo98} obtains that the scattering
length $a_{3\,3}^1 $ changes from $34.5~{\rm GeV}^{-3} $ at first
order, to $ 80.5~{\rm GeV}^{-3} $ at second order and to $ 81.4~{\rm
GeV}^{-3} $ at third order. Obviously, any attempt to unitarize this
amplitude based on the smallness of the second order term with respect
to the first order one, will {\it formally} reproduce HBChPT but it
will likely fail {\it numerically} to reproduce the corresponding
phase shifts even in the vicinity of the threshold region. Thus, to be
consistent with the calculation of Ref.~\cite{Mo98} the used
unitarization method should take into account the slow convergence
rate of HBChPT, and ensure, to start with, that it numerically
reproduces the scattering data around threshold.

The $\Delta$ resonance can be included as an explicit
degree of freedom  within HBChPT~\cite{ET97,HH97,MO99}. This
requires the introduction of new parameters into the Chiral
Lagrangian. On the other hand, the unitarization of the amplitude via
the Inverse Amplitude Method (IAM) does not introduce new parameters
and has recently been proposed~\cite{GP99}, but the second order terms
have been considered to be small. As a consequence, the low energy
constants turn out to be very different from those found within
HBChPT~\cite{Mo98} (see also  the discussion below). In particular,
their value $b_{19}=-23.13 $, implies a pion nucleon coupling constant
$ g_{\pi NN}= 19.23 $, a rather odd one, compared with the experimental
one $ g_{\pi NN}=13.4 \pm 0.1 $. In the present work we show how the
IAM should be modified in  order to simultaneously i)
implement exact unitarity , ii) comply with  HBChPT at threshold and
iii) describe the $\Delta $ resonance without having to introduce 
additional explicit parameters than those needed at  threshold.  

\section{ Partial Wave Amplitudes}

We rely heavily on the notation of Ref.~\cite{Mo98} and refer to that work 
for more details. The $\pi N $ scattering amplitude is given by 
\be 
T_{\pi N}^\pm =  
\bar u ( \krig{M} v + p' , \sigma' ) \Big[ A^\pm + B^\pm { \slashchar{q} + 
\slashchar{q}' \over 2} \Big] 
     u ( \krig{M} v + p , \sigma )
\ee
where $(\krig{M}v+p,q) $ and $(\krig{M}v+p',q')$ are the incoming and 
outgoing nucleon and pion momenta respectively. The bare nucleon mass 
$\krig{M}$ and the velocity 
$v^\mu $, ( $v^2=1$ ) determine the baryon off-shellness $p$ and $p'$ and 
$\sigma$ and $\sigma'$ are the spin indices. 
The supperscrit ``$ \pm $'' corresponds to 
the isospin decomposition $ T^{ba}=T^{+} \delta^{ba} +{\rm i} 
\epsilon^{bac} T^- \tau_c $, with $a$ and $b$ the incoming and outgoing pion 
isospin states in the cartesian basis respectively. $A^\pm $ and 
$B^\pm $ are scalars
depending on the Mandelstam variables $s,t$ and $u$ which for on-shell
pions and nucleons verify $ s+t+u=2(m^2 + M^2) $, with $m$ the physical pion 
mass and $M$ the physical nucleon mass. In the  CM frame, the partial wave 
amplitudes are written as (the subscript $\pm$ stands for $ j=l \pm 1/2 $), 
\begin{eqnarray}
 f_{l \, \pm}^\pm (s)= e^{{\rm i} \delta_{l \pm}^\pm (s)} 
{ {\rm sin}\delta_{l \pm}^\pm (s) \over q } = 
{1\over 16 \pi \sqrt{s}} & \Big( &(E+M)[ A_l^\pm (s)+ (\sqrt{s}-M)B_l^\pm (s)
\nonumber 
\\ & - & (E-M)[ A^\pm_{l\pm 1} (s) - (\sqrt{s}+M)B_{l\pm 1}^\pm (s)] \Big)
\end{eqnarray}
where, for elastic scattering, $q= |\vec q |$ is the CM momentum and 
$E=\sqrt{q^2 + M^2} $ and $ \omega =\sqrt{q^2+m^2} $ the CM nucleon and 
pion energies respectively. Finally, $ \sqrt{s}=E+\omega $ is the total CM 
energy and $ \delta_{l \pm}^\pm (s) $ are the corresponding phase-shifts. 
The projected amplitudes $A_l (s)$ and $B_l(s)$ are defined by the integrals
\be
\Big( A_l^\pm (s) , B_l^\pm (s) \Big)  =
 \int_{-1}^1 {\rm d} z \, \Big( A^\pm (s, z ), B^\pm (s, z ) \Big) P_l ( z )
\ee
with $ z= {\rm cos } \theta $ and $\theta$ the CM scattering angle. 
In addition, $A^\pm $ and $B^\pm $ can be written as 
\begin{eqnarray}
A^\pm &=& \Big( \alpha^\pm + {s-u \over 4} \beta^\pm \Big) \\
B^\pm &=& \Big(-M+{t\over 4M}\Big) \beta^\pm
\end{eqnarray}
with $ t=-2q^2(1-{\rm cos}\theta)$ and $u=M^2+m^2-2E\omega-2q^2 {\rm cos}
\theta$.
An expansion of both $\alpha^\pm $ and $\beta^\pm $ along the lines of HBChPT 
has been given in Ref.~\cite{Mo98} up to third order, i.e. including terms 
of order $e/f^2 $ (first order), $ e^2 /(f^2 M) $ (second order) and 
$e^3 /f^4 $ and $e^3/(f^2 M^2)$ (third order). There, however, in some cases 
the full nucleon mass dependence has been retained. This makes the discussion 
on orders a bit obscure\footnote{For instance, the $P_{33}$ scattering length
contains some corrections $m/M$ to all orders ( see eqs.(90-92) in that 
reference  ), instead of neglecting all pieces of order $1/(f^2 M^3)$, 
$ 1/(f^4 M )$ or higher as dictated by the expansion assumed in the Chiral 
Lagrangian.}. We have preferred to further expand any observable in terms of 
$1/M$ or $1/f^2$. As a consequence and in contrast to Ref.~\cite{Mo98}, there 
is no dependence of D,F, and higher partial waves on the low energy constants.
Thus, the low energy parameters are solely determined by S and P partial waves.
One thus gets the following expansion, 
\begin{equation}
 f_{l \, \pm}^\pm  =    f_{l \, \pm}^{(1) \, \pm}
                       +f_{l \, \pm}^{(2) \, \pm}
                       +f_{l \, \pm}^{(3) \, \pm}+ \cdots
\label{eq:fexpansion}
\end{equation}
where
\begin{eqnarray}
f_{l \, \pm}^{(1) \, \pm} &=&
{m\over f^2} t_{l \, \pm}^{(1,1) \, \pm} \({\omega \over m}\) \\
f_{l \, \pm}^{(2) \, \pm} &=&
{m\over f^2 M } t_{l \, \pm}^{(1,2) \, \pm} \({\omega \over m} \) \\
f_{l \, \pm}^{(3) \, \pm} &=&
{m^3 \over f^4 } t_{l \, \pm}^{(3,3) \, \pm} \( {\omega \over m} \)
+ {m^3 \over f^2 M^2 } t_{l \, \pm}^{(1,3) \, \pm} \({\omega\over m} \)
\label{eq:fexpansion2}
\end{eqnarray}
$t^{(n,m) \, \pm}_{l \, \pm}$  are dimensionless functions of the 
dimensionless variable $\omega/m$, independent of $f,M$ and $m$, whose 
analytical expressions are too
long to be displayed here. The convenience of the double superscript notation 
will be explained below; $n+1$ indicates the power in $1/f$ and $m$ the total 
order in the HBChPT counting. The unitarity condition $ {\rm Im}
f_{l \, \pm}^{-1} = -q $ becomes in perturbation theory 
\begin{eqnarray}
{\rm Im}\, t_{l \, \pm}^{(1,1) \, \pm} = 
{\rm Im}\, t_{l \, \pm}^{(1,2) \, \pm} &=& 
{\rm Im}\, t_{l \, \pm}^{(1,3) \, \pm} = 0 \\
{\rm Im}\, t_{l \, \pm}^{(3,3) \, \pm} &=& {q\over m}
| t_{l \, \pm}^{(1,1) \, \pm} |^2
\end{eqnarray}
The scattering lengths $ a_{l,\pm}^\pm $ and effective ranges $ b_{l,\pm}^\pm$
are defined by
\begin{equation}
{\rm Re} f_{l,\pm}^\pm = q^{2l} \Big( a_{l,\pm}^\pm + q^2 b_{l,\pm}^\pm + \cdots \Big)
\end{equation}
Obviously, the expansion of Eq.(\ref{eq:fexpansion}) for the amplitudes 
carries over to the threshold parameters, $a_{l,\pm}^\pm $ and $b_{l,\pm}^\pm 
$ (see below).

\section{ Re-fitting the Low Energy Constants} 

As we have said, in Ref.~\cite{Mo98}, a systematic expansion 
for the fixed-$t$ amplitudes $\alpha^{\pm}$  and $\beta^{\pm} $ was 
undertaken, but some higher order terms were retained in the partial wave 
amplitudes. This makes the discussion about orders a bit unclear, since 
different orders are mixed. In all our following considerations we further 
expand the 
amplitudes, and consequently the threshold and close to threshold parameters 
in the spirit of HBChPT. For the scattering lengths we have  
\begin{equation}
a_{2I \, 2J}^l  m^{2l+1} = {m\over f^2} \alpha^{(1,1)} + 
{m^2 \over f^2 M } \alpha^{(1,2)} + {m^3 \over f^2 M^2 } 
\alpha^{(1,3)} + {m^3 \over f^4  } \alpha^{(3,3)} + 
\end{equation}
and a similar expression for $ b_{2I \,2J}^l m^{2l+3} $. 
With this expansion we may reanalyze the fit of Ref.~\cite{Mo98}. For a better
comparison we do so for the {\it same} experimental data, and with the 
{\it same} input values of the parameters, $ M=939 \,{\rm MeV} $, $ m = 138\, 
{\rm MeV} $, $ f = 93 {\rm MeV} $ and $ g_A = 1.26 $. The result of the fit 
is presented in Table 1. As we see from the table the changes in the 
parameters are small, and are compatible within two standard deviations. 
This was to be expected since the difference between Mojzis's parameters 
and ours is of higher order. It is also noteworthy that the resulting 
$\chi^2 = 10.68 $ is mainly made out of $b_{0,+}^+ $ and $\sigma$ which 
contribute with 3.3 and 7.3 respectively to the total $\chi^2$. The bad 
result concerning the $\sigma$-term is substantiated by the findings of 
Refs.~\cite{BKM97,Mo98,FMS98}.  

After performing this re-fitting procedure we have found  instructive
to separate the contributions to the scattering lengths and effective ranges  
of the lowest partial waves, as done in Table 2. A clear distinctive pattern 
emerging from Table 2 is that 
the contribution of order $1/(f^2 M^2)$ is always rather small. Only in some 
cases, however, is the contribution of order $1/f^4 $ also small. This is so 
in the $P_{33}$ channel in particular. Incidentally, let us note that for 
this channel the smallness of the third order contribution is due 
to the smallness of both the $1/f^4 $ and the $1/(f^2 M)$  terms and does not
stem from a cancelation between large contributions. So, it seems  that 
close to threshold the $1/f^2$ expansion converges faster than the $1/M$ 
expansion. Actually, the terms in $1/f^2$ and $1/(f^2 M) $ are of comparable
importance. From here, it is clear that any unitarization method will only 
be consistent with the HBChPT approach, if it treats both the first and the 
second order as equally important.  

\begin{table}[t]
%\vspace{-0.3cm}
\begin{center}
\begin{tabular}{c|c|c}
        & Mojzis (Ref.~\protect{\cite{Mo98}})  & This Work \\    \hline 
        & Set {\bf I} & Set {\bf II} \\    \hline 
$ a_1$   & $-$2.60  $\pm$  0.03  & $-$2.72 $\pm$  0.03  \\
$a_2 $  & 1.40 $\pm$ 0.05   & 1.39$\pm$0.05   \\
$a_3 $  & $-$1.00 $\pm$ 0.06   & $-$1.11$\pm$0.06   \\
$a_5 $  & 3.30$\pm$0.05   &  3.3$\pm$0.07  \\
$\tilde b_1 + \tilde b_2 $  & 2.40$\pm$0.3   & 1.8$\pm$0.4  \\
$\tilde b_3 $  &    $-$2.8$\pm$0.6  & $-$3.4$\pm$0.7   \\
$\tilde b_6 $ &  1.4$\pm$0.3   & 1.9$\pm$0.3   \\
$ b_{16}- \tilde b_{15} $ &  6.1$\pm$0.6    & 6.1 $\pm$0.6    \\
$ b_{19}$ &   $-$2.4$\pm$0.4   & $-$2.4$\pm$0.4   \\
\end{tabular}
\end{center}
\caption[pepe]{\footnotesize Low energy constants obtained from
fitting the threshold parameters $a_{0,+}^\pm $, $b_{0,+}^\pm $, 
$a_{1,\pm}^\pm $ found in the present work and in
Ref.~\protect\cite{Mo98}, the Goldberger-Treiman discrepancy and the
nucleon $\sigma-$term to the HBChPT predictions. We take $ M=939 \,{\rm MeV} 
$, $ m = 138 \,{\rm MeV} $, $ f = 93\, {\rm MeV} $ and $ g_A = 1.26 $. 
Experimental data are taken from Ref.~\cite{KP80}. 
 The first column refers to the
work of Mojzis~\cite{Mo98} , where the fixed  $t$ amplitudes
$\alpha^\pm $ and $\beta^\pm $ have been expanded using HBChPT but
not the threshold parameters. In our case, we have further expanded
the threshold parameters (see text for details) and fitted to the same data. 
The dependence of the Goldberger-Treiman discrepancy and the nucleon 
$\sigma$-term on the fitted parameters can be found in Ref.~\cite{Mo98}. 
There is only one degree of freedom. Errors in any parameter have been 
calculated by changing the corresponding parameter until that the 
minimized $\chi^2 $ with respect to the remaining
parameters varies by one unit. The fit yields a 
$\chi_{\rm min}^2 $=10.68 which is basically made out of $b_{0,+}^+ $ and 
$\sigma$ which give 3.3 and 7.3 contribution to the total $\chi^2 
$ respectively. The value of the total $\chi^2 $ in Ref.~\cite{Mo98} is not 
given.} 
%\label{tab:HBCPT-th}
\end{table}
\begin{table}[t]
%\vspace{-0.3cm}
\begin{center}
\begin{tabular}{||c|c|c|c|c|c|c||}
  &  $1/f^2$  & $1/(f^2M) $ & $ 1/(f^2M^2)$ & $1/f^4$ & Total & 
Exp.(Ref.~\cite{KP80})  \\
\hline
$a_{3\,1}^0 $  
&  $-$0.63        & $-$0.08         &  +0.07       & $-$0.11    & $-$0.75  & $-$0.73 
$\pm$ 0.01  \\
$a_{1,1}^0 $  
&  1.27       & $-$0.36          &  $-$0.07        &  0.41    &  1.25  & 1.25 
$\pm$ 0.02 \\
$a_{3\,3}^1 $  
&  35.3       & 48.1          & $-$1.6        & $-$0.3    & 81.5  &  
81.5$\pm$0.8  \\
$a_{1\,3}^1 $  
&  $-$17.6       & 14.3          & $-$3.0        & $-$5.2    & $-$11.5  & $-$11.5 
$\pm$ 1.3 \\
$a_{3\,1}^1 $  
&  $-$17.6       & 11.8          & $-$1.7        & $-$9.6    & $-$17.1  & $-$17.2
$\pm$ 0.6 \\
$a_{1\,1}^1 $  
&  $-$70.6       & 86.9          & $-$3.6        & $-$43.1    & $-$30.4  &  
$-$30.4 $\pm$0.9   \\
\end{tabular}
\end{center}
\caption[pepe]{\footnotesize HBChPT, lowest partial S and P wave- scattering 
lengths, $a_{2I\,2J}^l $ (in ${\rm GeV}^{-2l-1} $ units) for $\pi N $ 
scattering in the in HBChPT, decomposed as a sum of terms of first order 
$ 1/f^2 $, second order, $ 1/(f^2 M ) $, and third order, $ 1/(f^2 M^2)$ and 
$ 1/(f^4) $. The sum of all the terms yields the total scattering length 
parameter. The experimental value obtained in Ref.~\cite{KP80} is also 
quoted.  }
%\label{tab:HBCPT-th}
\end{table}

\section{ Unitarization method for the $P_{33}$ channel}

Our unitarization method is based on assuming, as suggested by the
perturbative calculation, that the chiral expansion in terms $ 1/f^2 $, has 
a stronger convergence rate than the finite nucleon mass $1/M $ corrections. 
Such a hypothesis can only be supported by confrontation with  experimental 
data, and indeed there are some recent theoretical attempts\cite{BL99,Ge99} 
to define a relativistic power counting not requiring the heavy baryon idea
somehow retaking the spirit of older relativistic studies~\cite{gss88}. 
Indeed, we show that a slightly modified version of the well-known 
IAM approach turns out to describe the $P_{33}$ phase-shift 
satisfactorily, just using the low energy parameters determined at threshold, 
i.e., without re-fitting them in the unitarized case. 
The idea is to consider 
the expansion of the inverse amplitude in terms of $m^2 /f^2 $. The expansion
of the amplitude can be written as
\begin{equation}
f_{3\, 3}^1 (\omega, m , f , M) 
= {m\over f^2}   t^{(1)} ( \omega/m \, , \, m/M) +
                     {m^3\over f^4}   t^{(3)} ( \omega/m \, , \, m/M) + \dots
\end{equation}
where we have explicitly used that the functions $t^{(2n+1)}$ are
dimensionless, and hence depend  only on dimensionless variables, such as 
$\omega/m$ and $m/M$. For the sake of a lighter notation, the 
quantum numbers $l$  $I$ and $J$ have been purposely suppressed. 
Perturbative unitarity in this expansion requires at lowest order, 
\bea
{\rm Im}\, t^{(1)}(\omega/m , m/M) &=& 0 \\
{\rm Im}\, t^{(3)}(\omega/m , m/M) &=& {q\over m} |t^{(1)}(\omega/m , m/M)|^2
\eea
which in turn imply an infinite number of conditions in the $1/M$ expansion. 
The functions $t^{(1)}$ and $t^{(3)}$ are only known in a further $m/M$ 
expansion,
\begin{equation}
t^{(2n+1)} ( \omega/m  , m/M) =  t^{(2n+1,2n+1)} ( \omega/m  ) + {m\over M} \,
t^{(2n+1,2n+2)} ( \omega/m  ) + \, ({m\over M})^2 \, t^{(2n+1,2n+3)} 
( \omega/m  ) \, \dots
\end{equation}
yielding Eq.(\ref{eq:fexpansion}) and Eq.(\ref{eq:fexpansion2}) after 
a suitable isospin projection\footnote{ The relation between $ f_{2I,2J}^l $ 
and $ f_{l,\pm}^\pm $ is given by  $ f_{3,2l \pm 1}^l 
= f^+_{l,\pm} - f^-_{l,\pm}  $ and  $ f_{1,2l \pm 1}^l = f^+_{l,\pm} 
+2 f^-_{l,\pm}  $. }. For the inverse amplitude we get then 
\be
{1\over f_{3 \, 3}^1 ( s , m , f , M )} =
{f^2 \over m } {1\over t^{(1)}(\omega/m \,,\, m/M)}
 -m { t^{(3)} ( \omega/m \, , \, m/M) \over
[ t^{(1)} ( \omega/m \,   , \, m/M)]^2 } + \dots
\label{eq:invf2}
\ee
Obviously, expanding in $m/f^2$ may be justified provided these corrections
are small somewhere. As we have shown, for the $P_{33}$ channel, they are 
small precisely  at threshold. So in the threshold region the unitarization 
scheme will, approximately, reproduce the perturbative result. At the same 
time, unitarity is exactly implemented. However, the $1/M$ terms are not 
small corrections at threshold, so we refrain from further 
``expanding the denominator''. In this way we keep the first mass 
corrections as equally important. The state 
of the art in HBChPT calculations is such that only $ t^{(1,1)}$, 
$ t^{(1,2)}$, $ t^{(1,3)}$ and $ t^{(3,3)}$ are known. To ensure exact 
elastic unitarity we should keep only $t^{(1,1)}$ in the denominator of the 
second term in Eq.~(\ref{eq:invf2}) above. 
This induces a tiny error thanks to the 
smallness of $ t^{(3,3)}$ (correction $1/f^4$) in the $P_{33}$ channel. 
Of course, it would be highly desirable to compute at least $ t^{(3,4)}$ and 
$ t^{(3,5)}$ (orders $1/(f^4 M)$ and $1/(f^4 M^2)$ respectively)
in order to be able to keep also $ t^{(1,2)}$ in the denominator of the 
second term of the mentioned equation. After these remarks, we have  
\bea 
{1\over f_{3\,3}^1 ( s , m , f , M ) |_{\rm Unitarized}} &=&
{f^2 \over m } {1\over t^{(1,1)} ( \omega/m  ) + {m\over M} \,
t^{(1,2)} ( \omega/m  ) + \, ({m\over M})^2 \, t^{(1,3)} 
( \omega/m  ) } \nonumber
 \\ & &
 -m { t^{(3,3)} ( \omega/m ) \over
[ t^{(1,1)} ( \omega/m ) ]^2   } \label{eq:erajniam} 
\eea

At threshold, our formula yields a modified  scattering length 
\bea 
{1\over a_{\rm Unitarized}} &=&
{f^2 \over m } {1\over \alpha^{(1,1)} + {m\over M} \,
\alpha^{(1,2)} + \, ({m\over M})^2 \, \alpha^{(1,3)}  }  
 -m { \alpha^{(3,3)}  \over
[ \alpha^{(1,1)} ]^2   } \\
&=& {1\over a_{\rm HPChPT} -({m^3 \over f^4})\,  \alpha^{(3,3)} }  
- m { \alpha^{(3,3)}  \over [ \alpha^{(1,1)} ]^2   } 
\eea 
which on view of Table 2 yields, for the $P_{33}$ channel, the value 
$ a_{3 \,3}^1|_{\rm Unitarized}  = 80.2 {\rm GeV}^{-3} $, to be compared 
to $ a_{3 \,3}^1|_{\rm HBChPT}  = 81.0 {\rm GeV}^{-3} $, a compatible 
value with the experimental one. Notice that if
we {\it had} expanded the denominator considering the second 
order contribution ( $1/(f^2M)$ ) to be small we  would have obtained 
strictly the IAM method as used in Ref.~\cite{GP99},  
\be  
{1\over a_{\rm IAM}} =
{f^2 \over m } \left\{ {1\over \alpha^{(1,1)}}
- ({m\over M})\, { \alpha^{(1,2)} \over [ \alpha^{(1,1)}]^2 }
- ({m\over M})^2 \, { \alpha^{(1,3)} \over [\alpha^{(1,1)}]^2 }
+ ({m\over M})^2 \, { [\alpha^{(1,2)}]^2 \over [\alpha^{(1,1)}]^3 } \right\}
 -m { \alpha^{(3,1)}  \over  [ \alpha^{(1,1)} ]^2   } 
\ee
yielding $a_{3 \,3}^1 |_{\rm IAM}= 22.8 {\rm GeV}^{-3} $, a completely odd 
result. This explains why the low energy parameters recently found in 
Ref.~\cite{GP99} are so different from those found by Mojzis \cite{Mo98}. 
Actually, their value of $b_{19} = -23.13 $ would yield a pion-nucleon 
coupling constant $ g_{\pi NN} = 19.2$, completely out of question.

\section{ Numerical Results}

From our previous discussion it is clear that at threshold our unitarized 
amplitude  will reproduce very accurately and within error bars the HBChPT 
results and hence the experimental data. It is thus tempting to extend the 
$P_{33}$ phase shift  up to the resonance region and propagate the errors of 
the low energy parameters. We show in Fig.1 and Fig.2 the corresponding phase 
shifts as the outcome of our unitarization method, Eq.(~\ref{eq:erajniam}), 
and compare them to the experimental $\pi N $ data \cite{AS95}. We use both 
parameter sets given in Table 1. In Fig.2 we use the parameters determined 
in Ref.~\cite{Mo98} ( set {\bf I} ) and in Fig.1 those determined in the 
present work ( set {\bf II} ).

\begin{figure}
\begin{center}                                                                
\leavevmode
\epsfysize = 650pt
\makebox[0cm]{\epsfbox{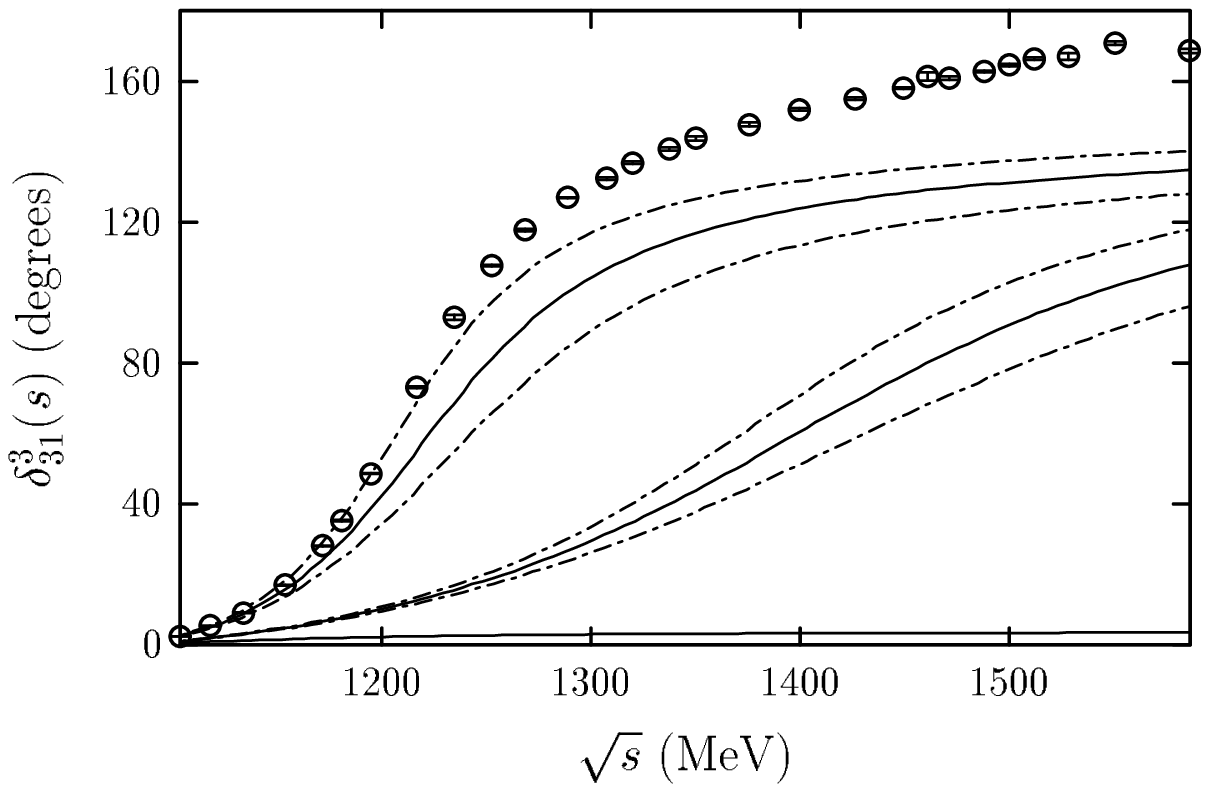}}
\end{center}
\vspace{-15.5cm}
\caption[pepe] {\footnotesize $P_{33}$ phase shifts as a function of
the total CM energy $\protect\sqrt s$. Experimental data are from
Ref.~\cite{AS95}. The upper solid curve is the result of our approach
(Eq.~\protect\ref{eq:erajniam}), with the corresponding propagated
error bars (dashed-dotted curve). The intermediate solid curve
represents the static limit prediction, $M\to \infty $, also with
error bands (dashed-dotted). The bottom solid line is the prediction
of the conventional IAM method of Ref.~\cite{GP99}. In all cases, the 
low energy parameters are those labeled as  set {\bf II} in Table 1. The
position and width of the resonance for the upper curve are $ M_\Delta
= 1267 \er{35}{25} \, {\rm MeV} $ and $ \Gamma_\Delta = 213
\er{84}{54} \, {\rm MeV}$ respectively.}
\end{figure}

As one can see from both figures, the prediction of Unitarized Heavy 
Baryon Chiral Perturbation Theory agrees rather well within errors with the 
data. It is fair to say, however, that for the parameters determined in 
Ref.~\cite{Mo98} the agreement seems much better, describing the data for 
larger CM energy values. We do not ascribe any particular significance to a 
very accurate agreement at larger energies since there is certainly more 
physics to be taken into account, like inelasticities. In both Fig.1 and 
Fig.2, the error bars reflect only the uncertainty in the low energy 
parameters and have been obtained by means of a Monte Carlo
gaussian sampling of the low energy parameters for any given CM energy
value. The location and width of the $\Delta$ resonance come out to 
be\footnote{ We obtain $M_\Delta $ from the condition $ \delta_{3,3}^1 
( M_\Delta^2 ) = \pi /2 $, and the width $\Gamma_\Delta$ from 
 $ 1 / \Gamma_\Delta = M_\Delta { {\rm d} \delta_{3,3}^1  (s) \over {\rm d} s}
|_{s=M_\Delta^2}$   } 
\begin{eqnarray}
M_\Delta &=& 1238 \er{22}{18} \, {\rm MeV}\, ({\rm set} \,{\bf I}),
 \quad 1267 \er{35}{25} \, {\rm MeV}\, ({\rm set}\, {\bf II}), 
 \quad  ( {\rm exp.}  \, 1232\pm 2 \, {\rm MeV} ) \nonumber 
\\
\Gamma_\Delta &=& 150 \er{43}{31} \,\,\,\, {\rm MeV}\,    
 ({\rm set}  \, {\bf I}) ,
 \quad  213 \er{84}{54} \,\,\,\, {\rm MeV}\, 
 ({\rm set} \,{\bf II}) ,
\quad ( {\rm exp.} \, 120\pm 10 \, {\rm MeV}) 
\label{eq:delta_mw}
\end{eqnarray}
in agreement within two standard deviations with the experimental numbers, 
although with much larger errors. 
Given the good quality of our description from threshold up to the $\Delta$ 
resonance, one might even re-fit the parameters of our unitarized amplitude, 
and hence reduce the errors of the low energy constants entering the 
definition of the amplitude  in the $P_{33} $ channel. In this way we would 
reduce errors quoted in Eq.(~\ref{eq:delta_mw}) as suggested in 
Ref.~\cite{FMS98}.     
\begin{figure}
\begin{center}                                                                
%\leavevmode
\epsfysize = 650pt
\makebox[0cm]{\epsfbox{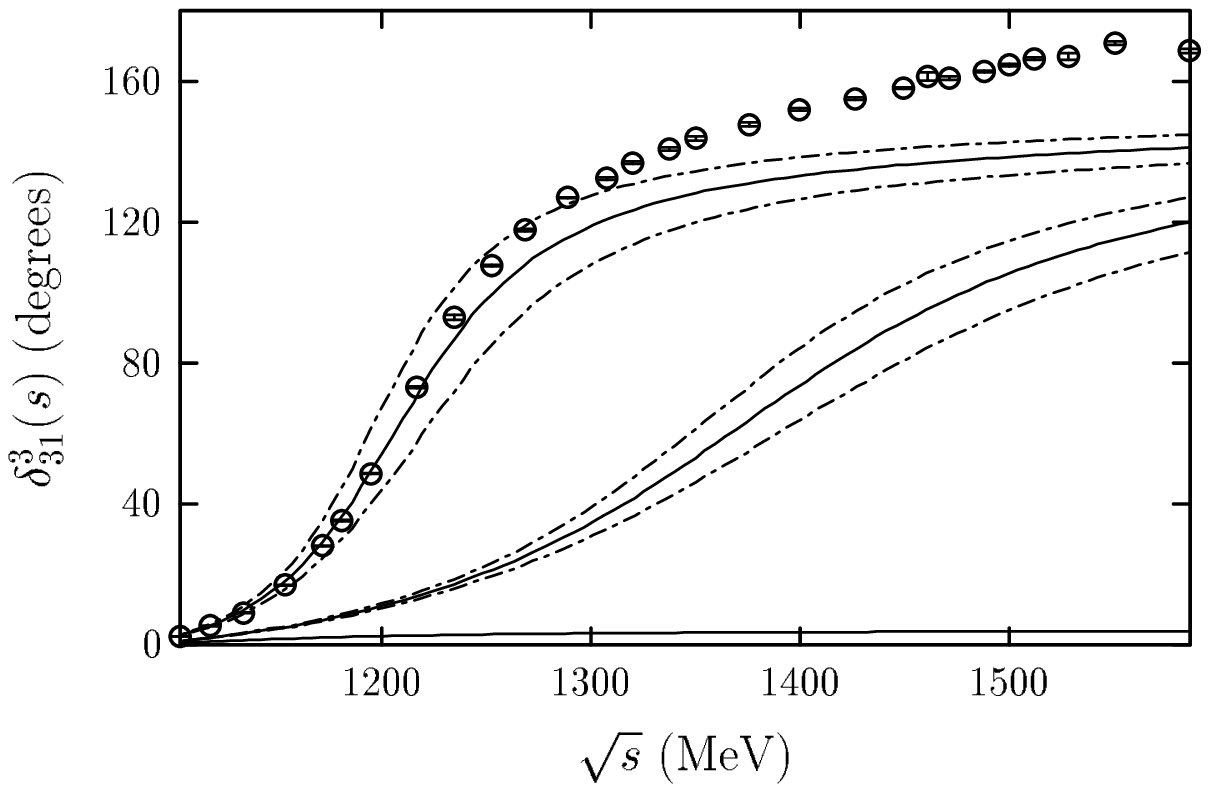}}
\end{center}
\vspace{-15.5cm}
\caption[pepe]
{\footnotesize The same as in Fig.1 but with the low energy 
parameters labeled as set {\bf I}. The position and 
width of the resonance for the upper curve are 
$ M_\Delta = 1238 \er{22}{18}\, 
{\rm MeV} $ and $ \Gamma_\Delta = 150 \er{43}{31} \, {\rm MeV}$ respectively.}
\end{figure}
To have an idea of the convergence rate of our calculation, we have
also depicted in both figures the prediction in the static limit ($ M
\to \infty $). As we see, there also appears a resonant behaviour,
but at higher energies $\sqrt{s} \sim 1450 {\rm MeV} $.  Thus, though
the bulk of the dynamics is contained in the static limit, the finite
mass corrections, particularly the $1/(f^2 M) $ contribution, are
important to achieve an accurate description\footnote{The $1/(f^2
M^2) $ correction turns out to be quite small. That was 
expected, since it neither provides a sizeable contribution at
threshold (unlike the $1/(f^2 M)$ correction ) nor it is responsible
for the restoration of unitarity (like the $1/f^4$ correction).}.

Finally we also show the results obtained within the conventional IAM
approach, see for details Ref.~\cite{GP99}, for both sets of parameters of
Table 1.  As we already anticipated, the description is extremely poor
mainly due to improper treatment of the $1/(f^2 M)$ corrections. Thus
within this framework, a fit to the data becomes possible only if huge
changes in the low energy parameters are allowed. For instance, the authors of 
that reference get $b_{16}- \tilde b_{15} = 68.9$, $b_{19} = -23.13$, 
$a_5 = 27.76$, $a_1 = 9.1$, $a_3 = -8.7 \cdots $ which strongly differ from 
the values of Table 1.

The situation in other $l_{2I2J}$ channels might be somehow different
and definitely deserves further study, though the results of the
present paper regarding the $P_{33}$ channel, will remain
unchanged. We anticipate, as can  be already seen from Table 2, that
the agreement of our unitarizated threshold parameters with those
stemming from HBChPT, would get a bit spoiled in some of the remaining
channels, since the $1/f^4$ is not a small contribution due to a,
perhaps accidental, but effective cancelation of the leading, $ 1/f^2
$, and the next to leading, $ 1/(f^2 M) $, order terms. The situation
might improve, however, if the low energy constants would be allowed
to vary as to provide a better description of the scattering data from
threshold up to the $\Delta$ resonance region.  A systematic and
detailed investigation of this topic will be presented elsewhere
\cite{EJ99bis}.

\section{Conclusions and Outlook}

We summarize our results. Heavy Baryon Chiral Perturbation Theory provides 
definite predictions for the $\pi N$ scattering amplitudes in the threshold 
region, but it violates exact unitarity if the perturbative expansion 
is truncated to some finite order. Hence it is unable to describe the 
$\Delta$ resonance in the $P_{33} $ channel. The analysis up to third order 
shows that the leading finite nucleon mass correction, which is second order, 
is of comparable size to the static approximation and in fact it dominates 
the corrections at threshold. This sets conditions on the unitarization method
suggesting an expansion in inverse powers of the weak pion decay constant
but without making the heavy baryon expansion. Such an idea is supported by 
recent theoretical attempts to redefine a relativistic chiral counting 
for baryons. Our unitarization method is a slight but important variant of 
the IAM and provides a {\it prediction} for the phase 
shifts in the $P_{33}$ channel, in terms of low energy parameters and their 
errors, as determined from  HBChPT. The agreement with experimental data is 
good within uncertainties. It is also clear that with our unitarization 
method, the combination of parameters which enter the $P_{33} $ amplitude, 
might be determined to a better accuracy than that obtained by only looking at 
threshold, since the central predicted values of the $P_{33}$  phase shift,
fall  mainly on top of the central experimental data.

%

%\newpage 

\section*{Acknowledgments}
This research was supported by DGES under contract PB98-1367 and by
the Junta de Andaluc\'\i a.

\end{document}